\documentclass{emulateapj}
\usepackage{graphicx}
\usepackage{natbib}

\shorttitle{A close pair of satellites in Centaurus A}
\shortauthors{Crnojevi\'c et al.}

\begin{document}


\title{Discovery of a close pair of faint dwarf galaxies in the halo of Centaurus A}


\author{D. Crnojevi\'c\altaffilmark{1}, D. J. Sand\altaffilmark{1}, N. Caldwell\altaffilmark{2}, P. Guhathakurta\altaffilmark{3}, B. McLeod\altaffilmark{2}, A. Seth\altaffilmark{4}, J. Simon\altaffilmark{5}, J. Strader\altaffilmark{6}, E. Toloba\altaffilmark{3}}





\begin{abstract}
As part of the Panoramic Imaging Survey of Centaurus and Sculptor
(PISCeS) we report the discovery of a pair of faint dwarf galaxies
(CenA-MM-Dw1 and CenA-MM-Dw2) at a projected distance of $\sim$90~kpc
from the nearby elliptical galaxy NGC~5128 (CenA).  We measure a tip
of the red giant branch distance to each dwarf, finding
$D$=3.63$\pm$0.41~Mpc for CenA-MM-Dw1 and $D$=3.60$\pm$0.41~Mpc for
CenA-MM-Dw2, both of which are consistent with the distance to
NGC~5128.  A qualitative analysis of the color magnitude diagrams
indicates stellar populations consisting of an old, metal-poor red
giant branch ($\gtrsim$12 Gyr, [Fe/H]$\sim$$-$1.7 to $-$1.9). In
addition, CenA-MM-Dw1 seems to host an intermediate-age population as
indicated by its candidate asymptotic giant branch stars. The derived
luminosities ($M_V=-10.9\pm0.3$ for CenA-MM-Dw1 and $-8.4\pm0.6$ for 
CenA-MM-Dw2) and half-light radii ($r_{h}=1.4\pm0.04$~kpc for 
CenA-MM-Dw1 and $0.36\pm0.08$~kpc for CenA-MM-Dw2) are consistent
with those of Local Group dwarfs. CenA-MM-Dw1's low central surface
brightness ($\mu_{V,0}=27.3\pm0.1$~mag/arcsec$^2$) places it among the
faintest and most extended M31 satellites. Most
intriguingly, CenA-MM-Dw1 and CenA-MM-Dw2 have a projected separation
of only 3 arcmin ($\sim3$~kpc): we are possibly observing the
first, faint satellite of a satellite in an external group of galaxies.

\end{abstract}

\keywords{galaxies: groups: individual (CenA) --- galaxies: halos --- galaxies: dwarf 
--- galaxies: photometry}

\altaffiltext{*}{This paper includes data gathered with the 6.5 meter Magellan Telescopes located at Las Campanas Observatory, Chile.}
\altaffiltext{1}{Texas Tech University, Physics Department, Box 41051, Lubbock, TX 79409-1051, USA}
\altaffiltext{2}{Harvard-Smithsonian Center for Astrophysics, Cambridge, MA 02138, USA}
\altaffiltext{3}{UCO/Lick Observatory, University of California, Santa Cruz, 1156 High Street, Santa Cruz, CA 95064, USA}
\altaffiltext{4}{Department of Physics and Astronomy, University of Utah, Salt Lake City, UT 84112, USA}
\altaffiltext{5}{Observatories of the Carnegie Institution for Science, 813 Santa Barbara Street, Pasadena, CA 91101, USA}
\altaffiltext{6}{Michigan State University, Department of Physics and Astronomy, East Lansing, MI 48824, USA}


\section{Introduction}

The relative number and astrophysical properties of dwarf galaxies
represent one of the major challenges to the widely accepted 
$\Lambda$+Cold Dark Matter ($\Lambda$CDM) cosmological model of
structure formation.  
For instance, the predicted number of DM haloes around a Milky Way (MW)-sized halo exceeds the 
observed number by at least an order of magnitude \citep[e.g.][]{klypin99,M99b}, and 
their central densities at a given mass are lower than
inferred from simulations \citep[e.g.][]{Boylan11,Boylan12}.

Understanding these challenges has largely been left to studies of the
Local Group \citep[LG; with exceptions, e.g.][]{Nierenberg13} where the faintest satellite 
galaxies can be discovered and studied in detail.  
However, a diversity of halo to halo substructure is expected due to varying accretion histories 
\citep{johnston08}  and inhomogeneous reionization \citep{busha10}; it is also possible that the 
LG is an outlier in its satellite properties.  A systematic study of the faint satellite populations 
of a large number of galaxies in a variety of environments is critical before the $\Lambda$CDM 
picture of galaxy formation can be confirmed.

Pioneering studies of faint satellite systems beyond the LG have begun.  The M81 group of galaxies 
(at $D$$\sim$3.6 Mpc) has been the subject of a resolved stellar population search for satellites 
utilizing deep ground based imaging and {\it Hubble Space Telescope} (HST) follow-up \citep{chiboucas09,chiboucas13}, 
finding satellites down to $M_V=-10$ and largely confirming the faint satellite deficit with respect to 
$\Lambda$CDM seen in the LG. Studies of unresolved low surface
brightness satellites have additionally
been undertaken for more distant groups \citep[e.g., M101,][]{merritt14}, where dwarf galaxy distance 
measurements are still necessary to produce conclusive results. 

We have begun the  Panoramic Imaging Survey of Centaurus and Sculptor (PISCeS) to significantly increase 
the sample of massive galaxy halos with a complete census of satellites down to $M_{V}$$\sim$$-$8~mag.  
PISCeS is focused on two nearby massive galaxies -- the spiral NGC253 and the elliptical NGC5128/CentaurusA 
(CenA), that reside in a loose group of galaxies (Sculptor) and in a rich group (CenA), respectively
\citep[see][for more details]{sand14}.  PISCeS will identify faint satellites and streams from 
resolved stellar light out to $R$$\sim$150~kpc in each system, allowing for direct
comparison with the MW, the M31 Pan-Andromeda Archaeological Survey \citep[PAndAS;][]{mcconnachie09} and 
M81 \citep{chiboucas13}. Out of the planned survey area of $\sim17$~deg$^2$,
we have already collected data over $\sim12$~deg$^2$ around CenA.

In this {\it Letter}, we present the first results of our survey around CenA: the
discovery of a close pair of faint dwarf galaxies, likely associated
with CenA. The dwarfs were immediately visible in our PISCeS survey data and are located near 
CenA's major axis in the north-east direction, at a projected
distance of $92$~kpc.
For the more luminous dwarf, a modest surface brightness enhancement can
be recognized in archival Digital Sky Survey images.

In \S \ref{sec:obs} we describe our data and reduction procedure, while in \S \ref{analysis}
we present the physical properties derived for the newly discovered dwarfs. In
\S \ref{concl} we discuss the possibility of these dwarfs constituting a pair of 
galaxies and draw our conclusions.


\section{Observations and data reduction} \label{sec:obs}

The data presented here were acquired on 2013 June 4 (UT), as part of the
PISCeS survey, with Megacam
\citep{mcleod06} at the Magellan Clay 6.5m telescope, which has a
$\sim$24'$\times$24' field of view and a binned pixel scale of
0\farcs16.  The seeing was excellent 
throughout the night, with a median of $\sim0.55-0.6"$, and the
conditions were photometric.  The final stacked images had a total exposure 
time of $5\times300$~sec for the $r$-band and $6\times300$~sec for the $g$-band. Initial 
data reduction (image detrending, astrometry and stacking) was performed
by the Smithsonian Astrophysical Observatory Telescope Data Center, 
using a code developed by M. Conroy, J. Roll and B. McLeod.

We perform point spread function (PSF)-fitting photometry on the final
stacked images, adopting the suite
of programs DAOPHOT and ALLFRAME \citep{stetson87, stetson94}. 
We construct a PSF for each band selecting $\sim450$ bright
stars across the image. We then fit the PSF to all objects 
$3\sigma$ above the background for each image, and compute the
coordinate transformations between filters with DAOMATCH/DAOMASTER 
\citep{stetson93}. Finally, we perform simultaneous photometry 
of all the objects detected in both filters in order
to obtain deeper catalogues. We only keep those objects satisfying the criteria
$\left|sharp\right|<3$ and $\chi<1.5$. The same night, several equatorial Sloan
Digital Sky Survey fields were observed at different airmasses to
obtain zeropoints, color terms and extinction coefficients, and thus
calibrate the instrumental magnitudes to the SDSS system.
Stars have been individually corrected for Galactic extinction 
\citep{schlafly11}, which has an average value of $E(B-V)$$\sim$0.15.

The photometric errors and incompleteness of our data have been computed
via artificial star experiments. We have injected $\sim$$10^6$
fake stars with a uniform distribution on each of the stacked images, divided 
into 20 experiments in order not to increase the stellar crowding artificially. 
The fake stars cover the whole relevant range in color-magnitude space, and additionally
reach $r\sim29$~mag ($\sim$2~mag fainter than the faintest real recovered stars).
The same PSF-fitting photometry procedure used for the real data has been 
performed on the fake stars. We find the overall (color-averaged) $50\%$ completeness 
limit to be $r\sim25.75$ and $g\sim26.75$.
Note that in the central $\sim$1~arcmin of CenA-MM-Dw1 the crowding is
higher than in the rest of the pointing, and the 
$50\%$ completeness limits correspond to $r\sim25.5$ and $g\sim26.5$.

Figure~\ref{cmds} shows the dereddened color magnitude diagrams (CMDs) 
for our newly found dwarfs, CenA-MM-Dw1 and CenA-MM-Dw2. The error bars (from the artificial 
star experiments) show the typical magnitude and color uncertainty as a function of 
$r$-band magnitude, and the $50\%$ completeness limits (for uncrowded regions) are drawn.


\section{Properties of CenA-MM-Dw1 and CenA-MM-Dw2} \label{analysis}

We begin this section by broadly discussing the stellar populations of our newly 
found dwarfs, and visualizing these new systems via red giant branch (RGB) stellar maps.  
We then discuss the distance, structure, metallicity and luminosity of this faint pair of CenA dwarfs.

\subsection{Stellar Populations and Red Giant Branch Map}

\begin{figure*}
 \centering
\includegraphics[width=10cm]{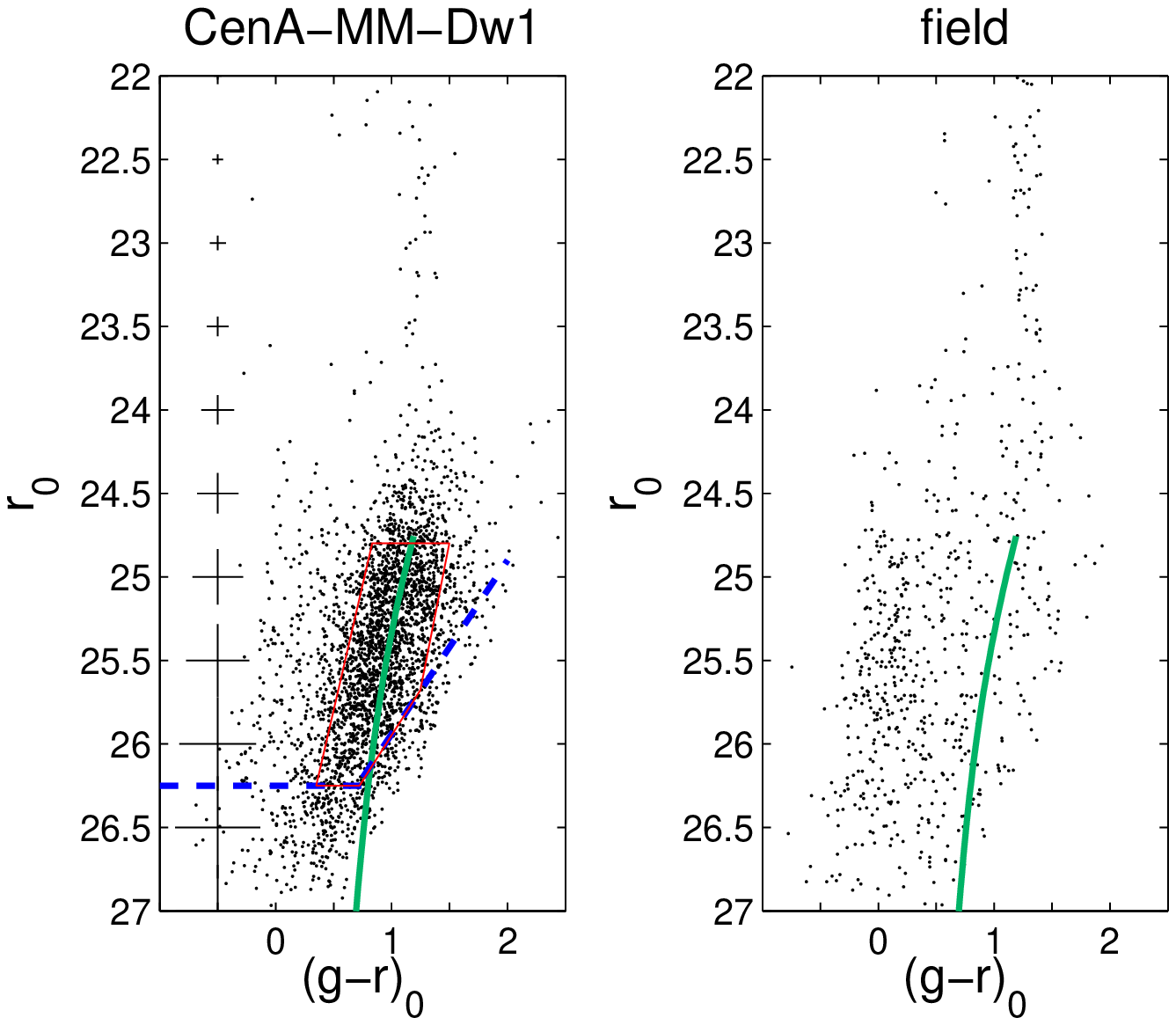}
\includegraphics[width=10cm]{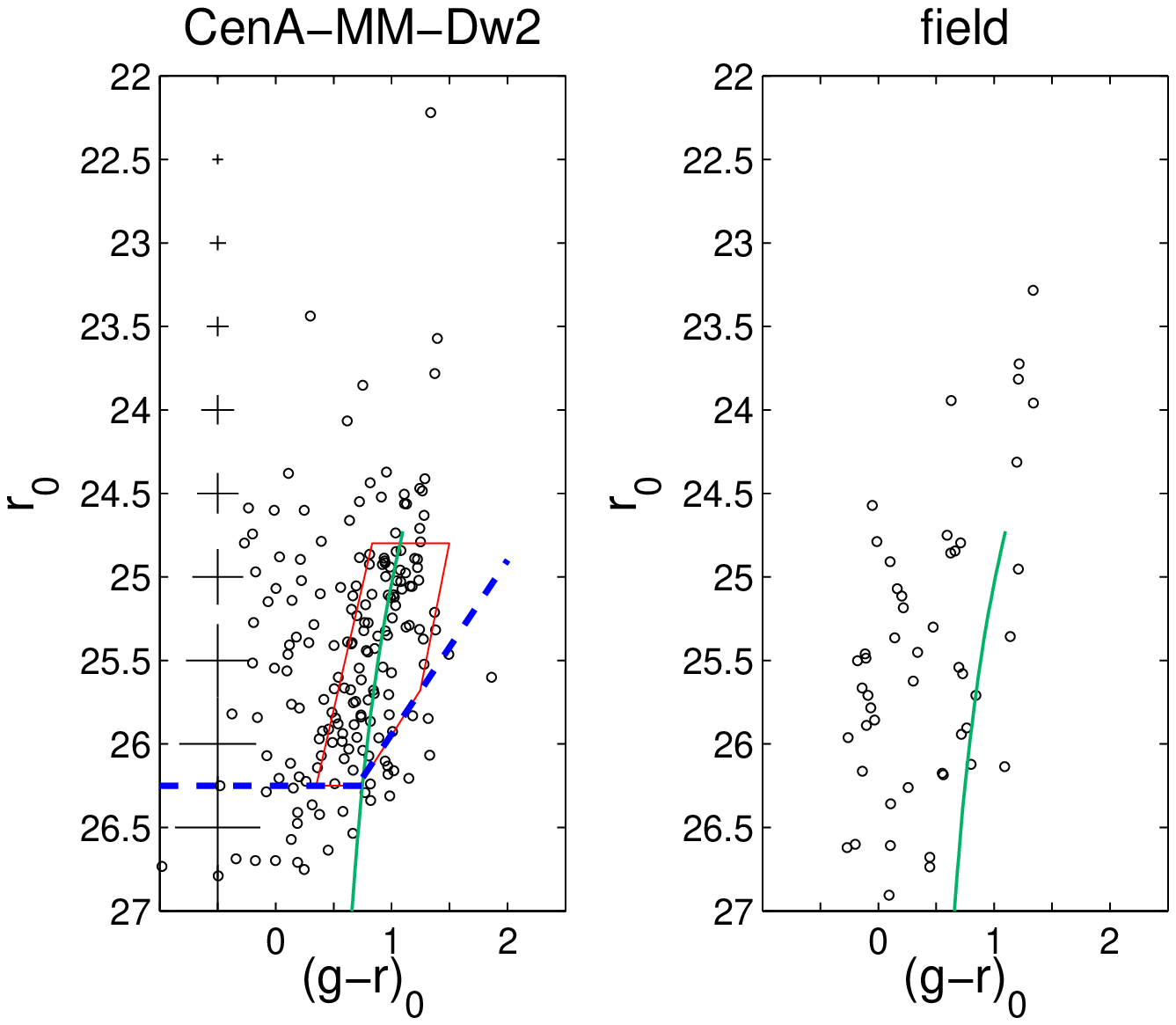}
\caption{De-reddened CMDs of the two newly
 discovered dwarf satellites (\emph{left panels}).
Stars within the half light radius are plotted.
 A background field CMD with the same area as
 the one selected for each dwarf is shown for comparison 
(\emph{right panels}). Overplotted on the clearly visible RGB
 for each dwarf is a Dartmouth isochrone shifted to the dwarf's 
 measured distance (\S~\ref{sec:distance}),
 with an age of 12~Gyr and metallicity [Fe/H]$=-1.7$ for
CenA-MM-Dw1 and $-2.0$ for CenA-MM-Dw2. The RGB selection 
box, utilized in Figure~\ref{rgb_maps}, is drawn in red.
The dashed lines indicate the $50\%$ completeness level while 
photometric errors stemming from artificial star tests are shown 
on the left side of each CMD.}
\label{cmds}
\end{figure*}

\begin{figure*}
 \centering
\includegraphics[width=7.cm]{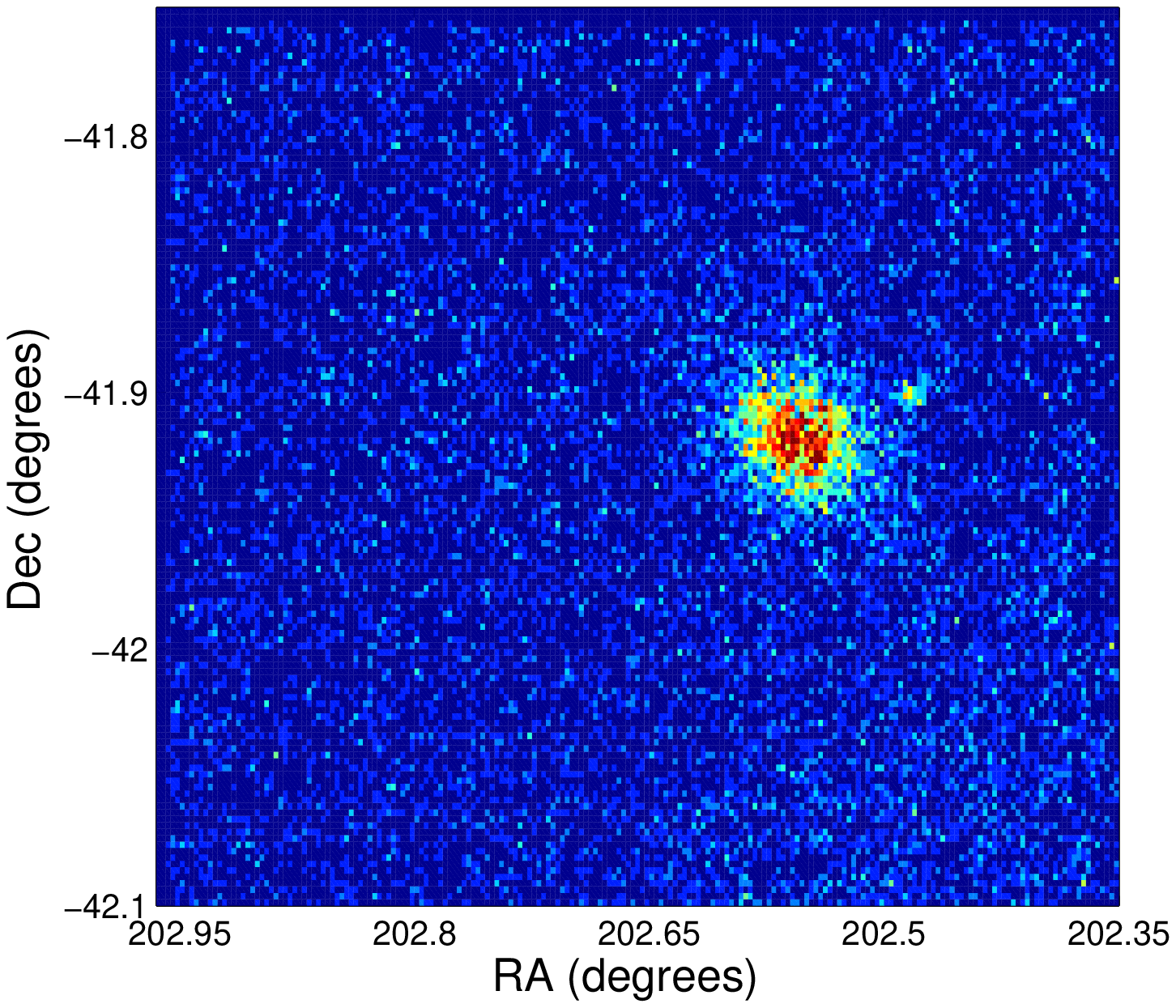}
\includegraphics[width=8.5cm]{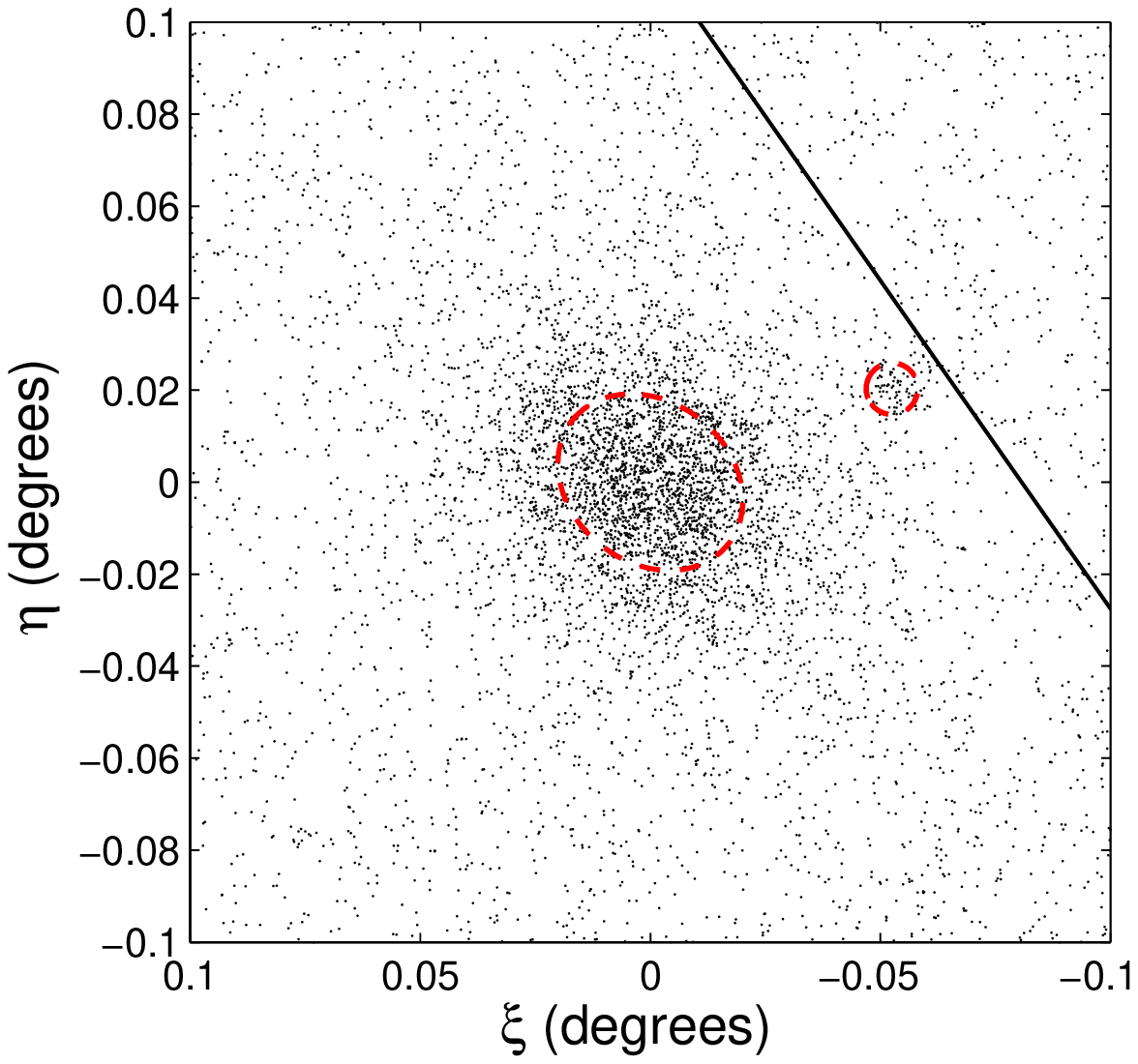}
\caption{\emph{Left panel}. Density map of all RGB stars
in the considered Magellan/Megacam pointing, including all stars 
drawn from the selection box illustrated in Figure~\ref{cmds}. The two newly
discovered dwarf satellites are clearly visible as overdensities, only $\sim$0.05~deg 
($\sim3$~kpc) apart in projection.
\emph{Right panel}. A zoomed-in view of the spatial distribution of RGB stars centered 
on CenA-MM-Dw1. For CenA-MM-Dw1 (CenA-MM-Dw2) 
we draw a red ellipse (circle) at the half light radius and with the 
$PA$ and $\epsilon$ reported in Table \ref{tab1}. 
The black line indicates the projection of CenA's major axis.}
\label{rgb_maps}
\end{figure*}

Figure~\ref{cmds} shows our dereddened CMDs where stars within 
the half light radius (calculated in \S~\ref{sec:struct}) are plotted.
The field CMDs (rescaled to the area of each dwarf) in the right panels are from 3 
rectangular regions with an area of $\sim0.1\times0.1$~deg$^2$ each,
two of which are located $\sim10$~arcmin closer/further away than our targets along CenA's 
major axis (see Figure~\ref{rgb_maps}), and the third one located
perpendicular to it, to the east of the two dwarfs.
This choice is driven by the necessity of taking into account a possible density
gradient and small-scale substructures in CenA's halo. The field CMDs show the
main contaminants, i.e. Galactic foreground (almost vertical) sequences at 
$(g-r)_0$$\sim$0.5 and $\sim$1.3, unresolved background galaxies centered
at $(g-r)_0$$\sim0$, and CenA RGB halo stars (sparsely populated at these
large projected galactocentric distances, $\sim$90-95~kpc). 

A prominent RGB is evident for both dwarfs. We plot the overall distribution of 
RGB stars in the Megacam field in Figure~\ref{rgb_maps}, utilizing all stars from the red 
selection box in the CMDs (Figure~\ref{cmds}). These predominantly old stars are clearly
clumped at the position of the dwarfs (Figure~\ref{rgb_maps}, right
panel). We cannot rule out the presence of younger, main sequence stars:
for CenA-MM-Dw2, a $\sim$1$\sigma$ excess of blue sources can be seen
at the center of the dwarf, but we do not consider this conclusive due to the strong
background galaxy contamination.
CenA-MM-Dw1's CMD shows an overdensity of
sources above the RGB with respect to foreground contaminants, which
may be luminous asymptotic giant branch stars. If real, they belong 
to an episode of star formation that occurred $\sim1-8$~Gyr ago.  A more definitive 
assessment of the dwarfs' star formation history will be forthcoming after planned 
HST observations of the pair.

\subsection{Distances to the new dwarfs} \label{sec:distance}

We derive the distance to the new dwarfs using the tip of the RGB (TRGB) 
method, which utilizes the brightest, metal poor RGB stars as a standard candle
\citep[e.g.][]{lee93, Salaris02, rizzi07}. We employ a Sobel edge detection
filter to identify a sharp transition in the $r$-band luminosity function, with a 
color cut of $0.8<(g-r)_0<1.2$ in order to 
minimize contamination from foreground Galactic stars (at red colors) and
unresolved background galaxies (bluer than the RGB).

We find $r_{0,TRGB}=24.79\pm0.22$ for CenA-MM-Dw1 and $r_{0,TRGB}=24.77\pm0.22$ 
for CenA-MM-Dw2, where the errors mainly depend on photometric uncertainties rather than the number of stars.
These values correspond to distance moduli of
$(m-M)_0=27.80\pm0.24$ and $(m-M)_0=27.78\pm0.24$, respectively, once adopting 
a theoretically calibrated TRGB absolute value of $M_r^{TRGB}=-3.01\pm0.1$ 
(as computed in \citealt{sand14} for SDSS bands). As a test, we derive
TRGB distances for the adopted field regions, i.e. for stars
belonging to CenA's outer halo. We obtain $(m-M)_0=27.85\pm0.28$, in 
excellent agreement with the average of literature values using several methodologies
\citep[$(m-M)_0=27.91\pm0.05$;][]{Harris10}. The derived distances put the 
two dwarfs at roughly the same distance as CenA, thus suggesting they are its
satellites. Follow-up HST data will be used to refine these distances 
and investigate the possibility that the two dwarfs form a physical pair
(see discussion in \S~\ref{concl}).

\subsection{Structural parameters and luminosities} \label{sec:struct}

To quantify the structure and luminosities of our new dwarf discoveries, we work with 
RGB stars from the selection boxes shown in Figure~\ref{cmds}. First, the dwarf centers are 
determined via an iterative process, computing the average of the stellar positions within circles of
decreasing radius.
From there, the position angle and ellipticity are measured using the method of moments for 
the RGB spatial distributions \citep[e.g.][]{mclaug94}. The final
values are reported in Table~\ref{tab1}. For CenA-MM-Dw2 we only obtain
upper limits on the ellipticity (and thus have a poorly determined position angle) because of the 
low number of available stars.

We derive radial density profiles by counting the number of RGB stars
in elliptical (circular) radii for CenA-MM-Dw1 (CenA-MM-Dw2). The
field level is estimated from the average number of RGB stars present in our
adopted field regions, and is entirely consistent with the number
density for CenA-MM-Dw1 beyond $\sim4.5$~arcmin. We subtract the field
level from the derived radial profiles and correct them for
incompleteness (including a correction for the excess crowding incompleteness 
within $\sim1$~arcmin of CenA-MM-Dw1).

\begin{figure*}
 \centering
\includegraphics[width=8.5cm]{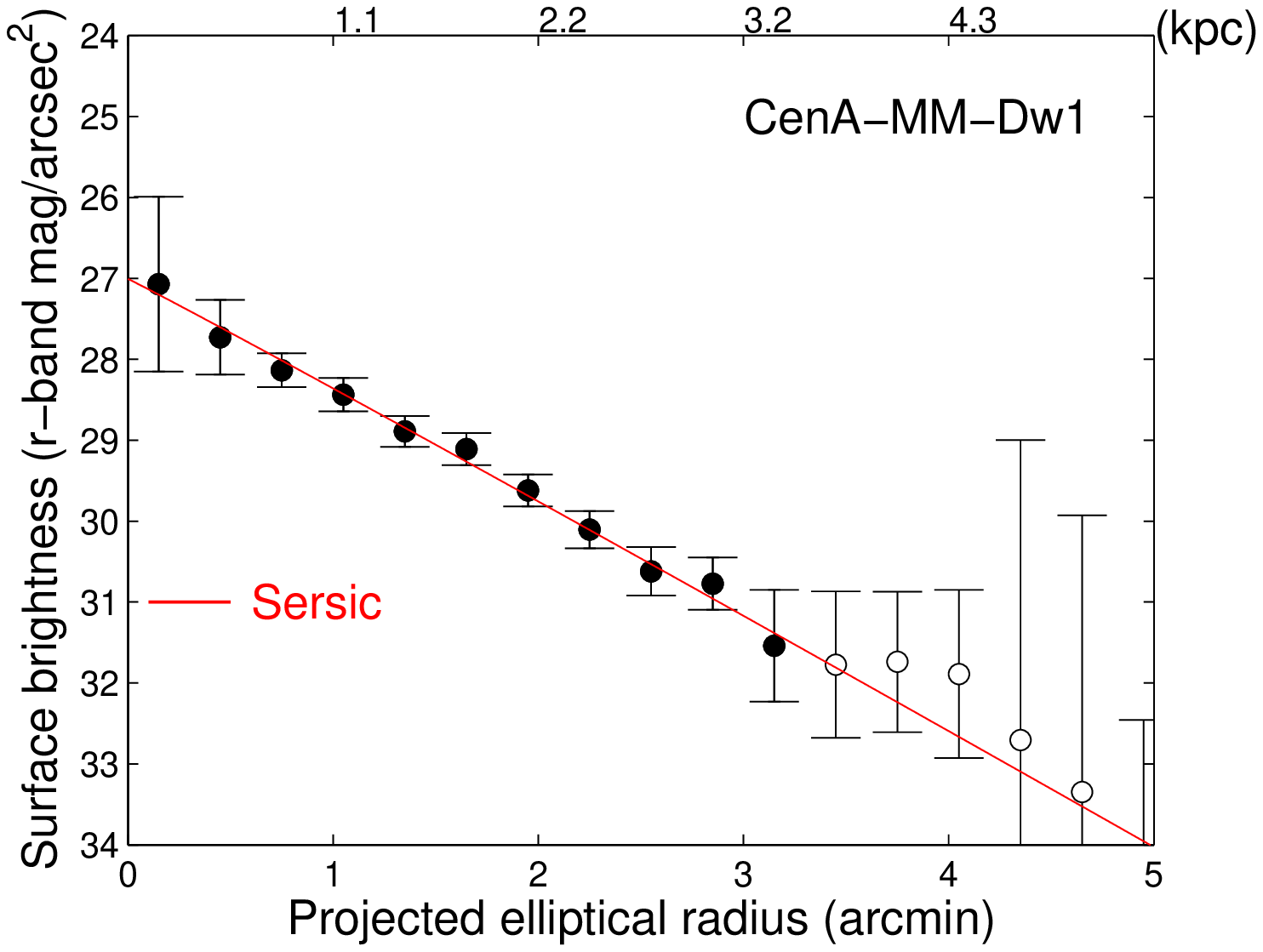}
\includegraphics[width=8.5cm]{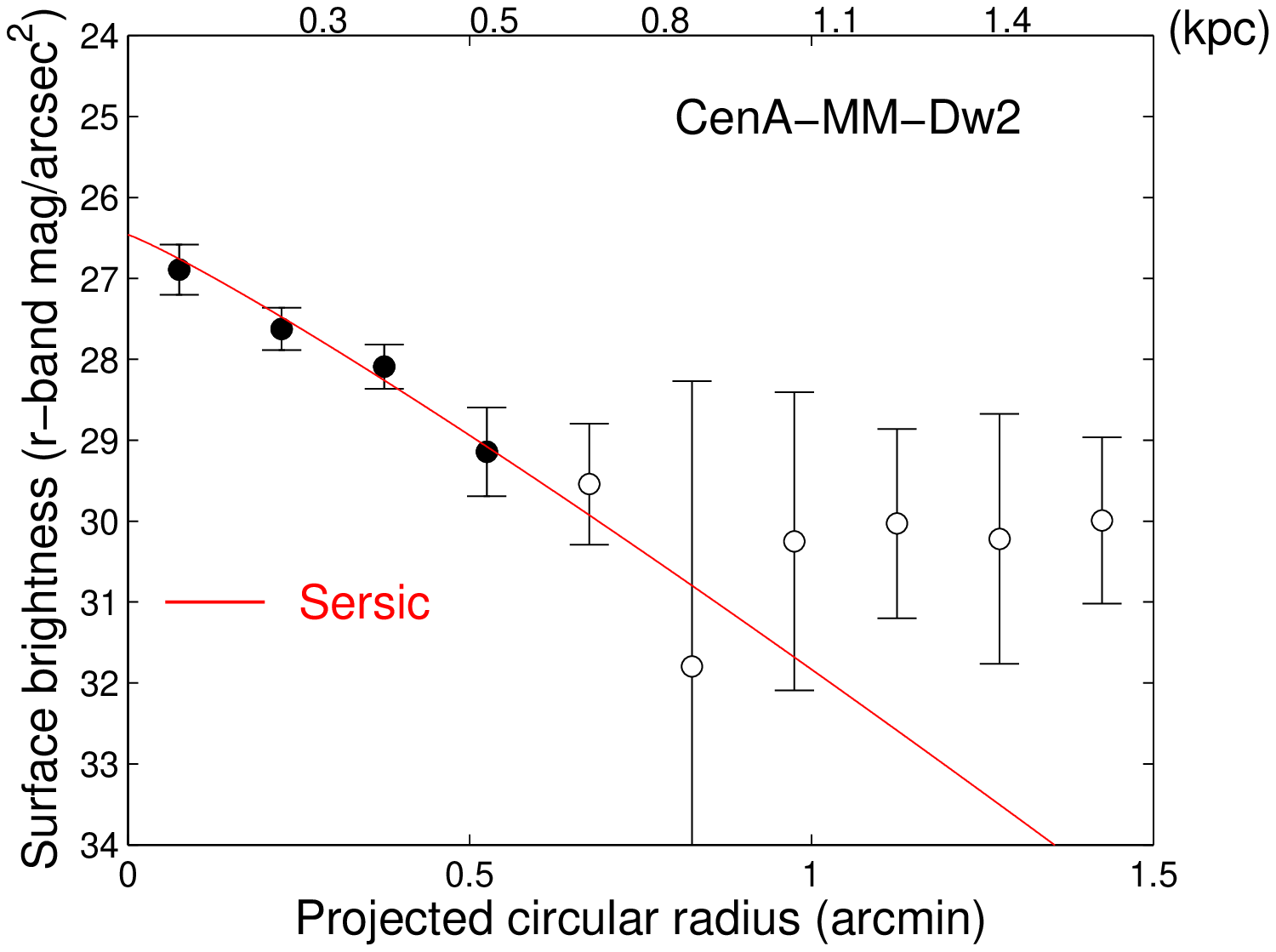}
\caption{Surface brightness profiles in $r$-band for the two dwarfs
as a function of elliptical (circular) radius for CenA-MM-Dw1 (CenA-MM-Dw2).
First, the number density profiles for RGB stars have been corrected for
incompleteness, and the field level has been subtracted from the profiles.
Subsequently, these have been converted into 
surface brightness by tagging them onto the integrated photometry within
the innermost $\sim0.2/0.1$~arcmin for CenA-MM-Dw1 and CenA-MM-Dw2, 
respectively. Error bars are Poissonian. 
CenA-MM-Dw2 can be recognized in CenA-MM-Dw1's profile as
an overdensity at $\sim$4~arcmin, while CenA-MM-Dw1 starts to dominate CenA-MM-Dw2's
profile beyond $\sim0.75$~arcmin. Filled symbols indicate the datapoints included in 
the Sersic fit, and the best-fitting Sersic profiles are overplotted,
being consistent with exponential profiles (see Tab.~\ref{tab1}).
The upper $x$-axes report galactocentric distances in physical units. 
}
\label{sbprofs}
\end{figure*}

Given the faint nature of these dwarfs, 
resolved RGB stars are necessary to trace their radial profiles as
opposed to integrated light. However, we adopt the latter to
convert the observed number density into proper surface brightness values.
This can be done by performing integrated aperture photometry in the
central regions of the dwarfs, and scaling the number density profile
to the central surface brightness values (e.g., \citealt{barker09,
  bernard12, crnojevic14}). The zeropoint
between the two profiles is computed from their overlapping region,
i.e. the innermost $\sim0.2/0.1$~arcmin for CenA-MM-Dw1 and
CenA-MM-Dw2, respectively. The integrated light photometry is performed 
by summing up the flux within these apertures, subtracting the median
sky level and applying zeropoint, color terms and airmass corrections
(Sect. \ref{sec:obs}).
The composite surface brightness profiles are shown in
Figure~\ref{sbprofs}. Note that we reach a remarkable surface brightness 
of $\mu_{r}\sim31$ mag/arcsec$^2$ (before field subtraction).

S\'ersic models are fit via least squares minimization 
to the composite surface brightness profiles (Figure~\ref{sbprofs}),  
for datapoints within $\sim3$~arcmin for CenA-MM-Dw1 and 
$\sim0.5$~arcmin for CenA-MM-Dw2. Beyond these radii, the profile of
each dwarf is contaminated by the light of the other dwarf
(CenA-MM-Dw2 can be recognized as an enhancement in CenA-MM-Dw1's
profile at $\sim4$~arcmin).
The resulting parameters of the Sersic profiles are reported in
Table~\ref{tab1}, and are consistent with being exponential profiles.

The uncontaminated profiles trace out to $\sim2$ times the derived 
half-light radii for the dwarfs.
We derive their luminosities by integrating the 
best-fit S\'ersic profiles, and obtain $M_r=-11.2\pm0.3$ for
CenA-MM-Dw1 and $M_r=-8.9\pm0.6$ for CenA-MM-Dw2
(uncertainties from error propagation). 
Using the transformation between SDSS and
Johnson-Cousins filters reported by \citet{jester05}, the 
$V$-band total magnitudes of the dwarfs
are $M_V=-10.9\pm0.3$ for CenA-MM-Dw1 and $-8.4\pm0.6$ for 
CenA-MM-Dw2.

\subsection{Metallicities}

The two dwarfs host relatively metal-poor stellar populations, as can
be seen from the isochrones overplotted on the CMDs in
Figure~\ref{cmds}.  To quantify this, we use the standard method for
computing photometric metallicities \citep[e.g.][]{crnojevic10}:
we interpolate between solar-scaled isochrones with a fixed age of 12~Gyr and [Fe/H] 
varying from $-2.5$ to $-1.1$, adopting the Dartmouth stellar 
evolutionary database \citep{dotter08}.  A metallicity value is obtained for each individual RGB star 
with magnitude $r_0<25.5$.  The mean values we deduce are [Fe/H]$\sim-1.7$ for CenA-MM-Dw1 and
[Fe/H]$\sim-1.9$ for CenA-MM-Dw2 (as plotted in Figure~\ref{cmds}).
The spread of the RGB is consistent with photometric errors, however we cannot 
exclude an intrinsic range of metallicities for 
CenA-MM-Dw1 in particular.  Follow-up data will be crucial to assess
the metallicity content of these dwarfs more precisely.


\section{Discussion and Conclusions} \label{concl}

We have presented the discovery of a faint pair of (likely) satellites of
the elliptical galaxy CenA, discovered within our PISCeS survey. 
CenA-MM-Dw1 and CenA-MM-Dw2 are located at a projected 
distance of 92~kpc from the center of their parent galaxy and lie at its approximate distance
($\sim3.6\pm0.4$~Mpc), although the uncertainties call for deeper
photometric follow-up. Both dwarfs contain predominantly old ($\sim$12~Gyr) and
metal-poor stellar populations ([Fe/H]$\sim-1.9/-1.7$), while
CenA-MM-Dw1 likely also contains intermediate age asymptotic giant branch stars.  
Neither CenA-MM-Dw1 or CenA-MM-Dw2 are detected in the HI Parkes All Sky Survey 
\citep[HIPASS;][]{Barnes01}, with 3-$\sigma$ upper mass limits of $M_{HI}\lesssim 4\times10^{7} M_{\odot}$.

\begin{figure*}
 \centering
\includegraphics[width=8.5cm]{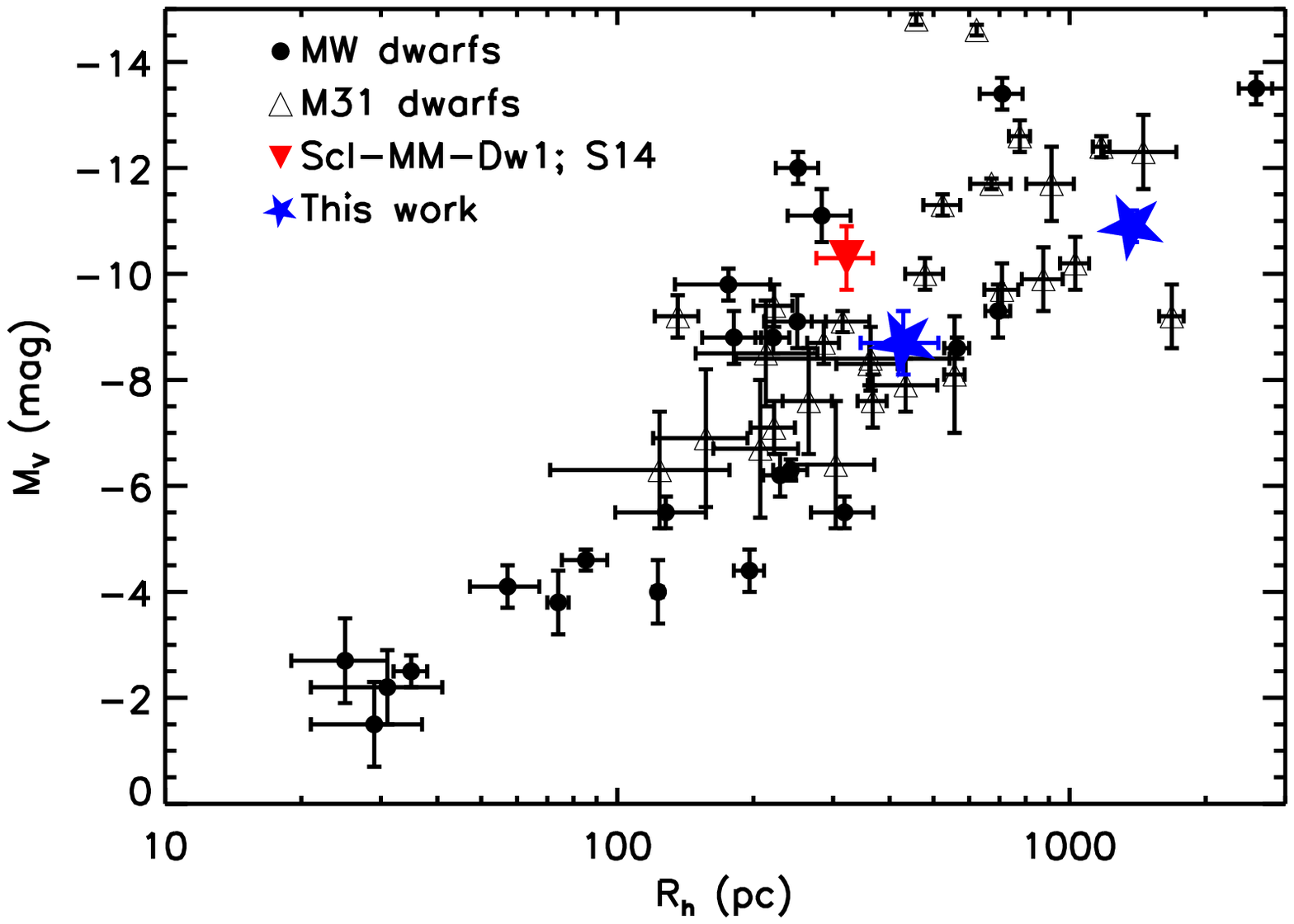}
\includegraphics[width=8.5cm]{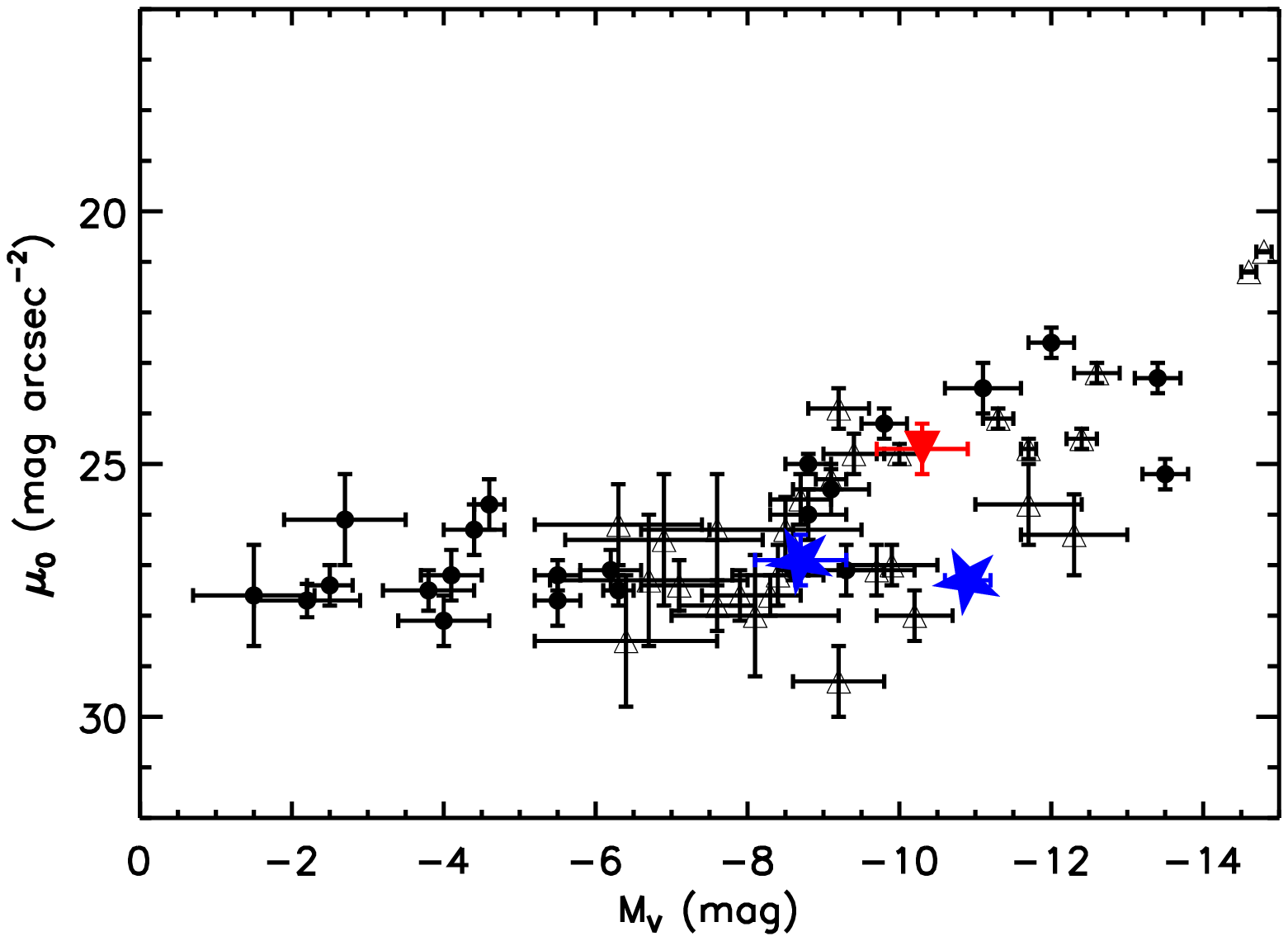}
\caption{\emph{Left panel}: Absolute $V$-band magnitude
as a function of half-light radius for MW/M31 dwarf galaxies 
(black points/triangles, from \citet{mcconnachie12} or \citet{Sand12}), 
CenA-MM-Dw1 and CenA-MM-Dw2 (blue stars), and Scl-MM-Dw1 
(red inverted triangle, \citealt{sand14}). 
\emph{Right panel}: Central $V$-band surface brightness as
a function of absolute magnitude. 
CenA-MM-Dw1's properties place it among M31 companions with 
the lowest central surface brightnesses and largest half-light radii.}
\label{rhmv}
\end{figure*}

The derived structural and luminosity parameters of
CenA-MM-Dw1 and CenA-MM-Dw2  (Table~\ref{tab1}) place them within the
main locus defined by MW and M31 dwarfs with similar properties (Figure \ref{rhmv}).
Of some note is the low central surface brightness of 
CenA-MM-Dw1 ($\mu_{V,0}=27.3$~mag/arcsec$^2$), which lies 
at the edge of the M31 satellites distribution in a surface brightness
versus luminosity plot.
Its properties are comparable to a few M31 satellites \citep[AndXXIII, AndXIX, LacI and
CasIII;][]{mcconnachie12, Martin13b} with large half light radii
($\gtrsim1$~kpc) and unusually low central surface brightness
($\mu_{V,0}\gtrsim26-28$~mag/arcsec$^2$). 
No MW satellite shows such a low central surface
brightness and large half light radius at the same time.
It has been suggested that M31
companions with $M_V>-9$ have lower surface brightness and are more 
extended with respect to MW companions with similar luminosity
\citep{kalirai10}, although the inclusion of the latest results for
the M31 subgroup have substantially decreased this difference
\citep{tollerud12}. 

The new dwarfs' distances are consistent 
with each other, and their very small angular separation
(3~arcmin or 3.1~kpc at the distance of the dwarfs)
suggests we are looking at the first, faint satellite of a satellite 
in an external group of galaxies.
According to simulations, groups of dwarfs infalling into the
potential well of a giant host are common \citep{donghia08},
although the predicted pairs of satellites (with comparable
luminosities) are fewer than those observed in
the LG \citep[e.g.][]{fattahi13}.

Examples of dwarf associations are present within the LG:
the Magellanic Clouds,
NGC147/NGC185 (\citealt{fattahi13, crnojevic14}, but see
also \citealt{watkins13}), Leo IV/Leo V/Crater 
\citep[][but see \citealt{Sandleoiv,Sand12,Jin12}]{leov, leoivleov, belokurov14}, and for 
all these pairs a common infall to the LG has been suggested
\citep{Evslin14}. Note that all the galaxies within these 
pairs display similar luminosities,
while CenA-MM-Dw1 is $\sim2.5$~mag brighter than CenA-MM-Dw2.
In some cases, stellar streams/overdensities with small angular separation are
interpreted as dynamically associated structures, or even remnants of 
infalling groups onto the MW's halo, where the
most massive of the dwarfs has already been disrupted
\citep[][and references therein]{Belokurov09, Deason14a}.
\citet{Deason14a} suggest that $30\%$ of faint dwarfs are likely to 
have fallen onto the MW as satellites
of more massive dwarfs.

For some of the cited pairs, tidal perturbances/streams have been
observed, even though their origin is still under
discussion. Intriguingly, CenA-MM-Dw2 lies only $\sim3$~arcmin in 
projection away from its more massive companion and does not display clear 
signs of distortion, however the upper limit for its ellipticity 
($\epsilon=0.67$) is high, and possible substructures 
may be hiding at fainter magnitudes.
We compute CenA-MM-Dw1's tidal radius from a King profile, obtaining
$6.3\pm0.9$~kpc, while its Jacobi radius is $2.2\pm0.3-4.7\pm0.5$~kpc
(assuming $(M/L)_V=10-100$, a projected radius from CenA of 92~kpc and
a CenA mass of $0.5\times10^{12} M_{\odot}$, \citealt{kara07}),
thus CenA-MM-Dw2 may be within its gravitational influence.
The CMDs derived from deeper HST imaging will allow for an improved
rejection of background galaxies, and thus a more accurate investigation of
CenA-MM-Dw2's structural parameters and possible tidal distortions.  
If this speculation is confirmed, we might be looking at the last
moments of CenA-MM-Dw2 before it is accreted by its more massive companion CenA-MM-Dw1.


\begin{deluxetable*}{lcc}
\tablecolumns{3}
  \tablecaption{Properties of the newly discovered dwarfs.}

\tablehead{
\colhead{Parameter}  & \colhead{CenA-MM-Dw1} &\colhead{CenA-MM-Dw2 } \\
}\\
  
\startdata
RA (h:m:s) & 13:30:14.26$\pm2$'' & 13:29:57.34$\pm2$'' \\
Dec (d:m:s) & $-41$:53:35.8$\pm10$'' & $-41$:52:22.6$\pm10$''\\
$(m-M)_0$ (mag) & $27.80\pm0.24$ & $27.78\pm0.24$ \\
D (Mpc) & $3.63\pm0.41$& $3.60\pm0.41$\\
$\epsilon$ & $0.19\pm0.01$& $<$0.67\tablenotemark{a} \\
$PA$ (N to E; $^o$) & $51.3\pm1.1$ & Unconstrained\tablenotemark{a}\\
$\mu_{r, h}$ (mag/arcsec$^2$) & $28.8\pm0.1$ & $28.1\pm0.5$ \\ 
$r_{h}$ (arcmin) & $1.30\pm0.04$ & $0.34\pm0.08$ \\ 
$r_{h}$ (kpc) & $1.4\pm0.04$ & $0.36\pm0.08$ \\ 
$n$ (Sersic index)  & $0.98\pm0.10$ & $0.90\pm0.44$ \\ 
$\mu_{V,0}$ (mag/arcsec$^2$) & $27.3\pm0.1$ & $26.5\pm0.5$ \\ 
$M_V$ (mag) &$-10.9\pm0.3$ &$-8.4\pm0.6$ \\
$L_*$ ($10^6 L_\odot$) &$2\pm0.5$ &$0.2\pm0.1$ \\
\enddata
\tablenotetext{a}{Only upper limits on the ellipticity, $\epsilon$, of 
CenA-MM-Dw2 were measurable, thus leaving the $PA$ unconstrained.}\label{tab1}

\end{deluxetable*}


\section*{Acknowledgements}

We warmly thank Maureen Conroy, John Roll and Sean Moran for their prolonged
efforts and help related to Megacam.
DC wishes to kindly thank the hospitality of the Mullard Space Science
Laboratory, University College of London, where part
of this work has been carried out.  DJS, JS and PG acknowledge support from NSF grant AST-1412504;
PG acknowledges additional support from NSF grant AST-1010039.
This paper uses data products produced by the OIR Telescope
Data Center, supported by the Smithsonian Astrophysical
Observatory.


\clearpage


\end{document}